\documentclass{article}
\usepackage{graphicx}  
\usepackage{amsmath}   
\usepackage[compress]{cite}
\usepackage{amssymb}   
\usepackage{bm} 
\usepackage{dcolumn}
\usepackage{color}
\usepackage{mathrsfs}
\usepackage{amsfonts}
\usepackage{varioref}
\usepackage{textcomp}
\usepackage{xcolor}
\RequirePackage[colorlinks,citecolor=blue,urlcolor=magenta,linkcolor=blue]{hyperref}
\allowdisplaybreaks
\def\gr{general relativity}

\labelformat{section}{Section #1} 
\labelformat{subsection}{Section #1} 
\labelformat{subsubsection}{Section #1}
\labelformat{subsubsubsection}{Section #1}
\labelformat{equation}{Eq.~(#1)} 
\labelformat{figure}{Fig.~#1} 
\labelformat{subfigure}{Fig.~\thefigure#1} 
\labelformat{table}{Table~#1} 
\labelformat{appendix}{Appendix #1}
\title{Non-trivial time crystal-like ground state for gravitational perturbation in quadratic gravity}
\author{Sumanta Chakraborty\footnote{sumantac.physics@gmail.com}$~^{1}$ and Subir Ghosh\footnote{subirghosh20@gmail.com}$~^{2}$\\
{$^{1}$\small{School of Physical Sciences, Indian Association for the Cultivation of Science, Kolkata-700032, India}}\\
{$^{2}$\small{Physics and Applied Mathematics Unit, Indian Statistical Institute, Kolkata-700108, India}}}
\begin{document}
  
\maketitle

\begin{abstract}
We consider gravitational perturbation around maximally symmetric background of a theory of gravity involving quadratic curvature correction. This leads to a decoupled model of the standard transverse, traceless graviton mode and an additional scalar degree of freedom. Evolution of the latter is governed by a Lagrangian, which depends on higher derivative terms, inherited from the quadratic curvature correction in the action. It turns out that stability of the gravitational theory allows the scalar sector to choose between possible lowest energy states (or, ground states), that can sustain periodic form of inhomogeneity, either in space or in time. The latter, with a restricted form of invariance in time translation, is possibly a variant form of the time crystal. Existence of this novel ground state, depicting a time crystal, is solely related to the presence of higher curvature terms in the action. In particular, this non-trivial condensate-like ground state describe the spacetime fabric itself by introducing fundamental scales in the theory. Possible implications are also discussed.
\end{abstract}
\section{Introduction} 

Recent detection of gravitational waves from the collisions between binary black holes and neutron stars as well as direct measurement of the black hole shadow have tested general relativity to an unprecedented accuracy \cite{Abbott:2016nmj,TheLIGOScientific:2016src,Abbott:2017vtc,Abbott:2016blz,Akiyama:2019cqa,Akiyama:2019bqs,Akiyama:2019eap}. This has further cemented the place of general relativity as the theory describing gravitational interaction. Besides such astonishing success of general relativity, it also suffers from several shortcomings. In the small length scale, the appearance of black hole singularity breaks the predictability of Einstein's equations and hence renders the theory useless. In an identical manner, at large length scale one has to invoke some exotic matter, e.g., dark energy, for the theory to be consistent with recent observations \cite{Aghanim:2019ame,Aghanim:2018oex,Clemson:2011an,Lewis:2002ah,Aghanim:2018eyx,Ade:2015xua,Bamba:2012cp,Copeland:2006wr,Peebles:2002gy,Padmanabhan:2002ji}. Despite several attempts, till date, there are no satisfactory resolution to the black hole singularity, but in the cosmological context, certain modifications over and above \gr\ may lead to late time cosmic acceleration without invoking any exotic matter \cite{Elizalde:2004mq,Garcia-Aspeitia:2018fvw,Bloomfield:2012ff}. Such modifications can not be arbitrary, since the resulting field equations must admit Schwarzschild like solutions, so that local physics remains identical to that of general relativity. Among various such modifications, $f(R)$ gravity \cite{Sotiriou:2008rp,Nojiri:2007as,Nojiri:2003ft,Nojiri:2006be,DeFelice:2010aj,Nojiri:2010wj,Chakraborty:2016ydo,Chakraborty:2014xla,Chakraborty:2015bja,Chakraborty:2020bne}, Lanczos-Lovelock models of gravity \cite{Chakraborty:2015wma,Padmanabhan:2013xyr,Chakraborty:2018dvi,Chakraborty:2017zep,Chakraborty:2015taq} as well as a certain class of scalar-tensor theories of gravity, known as Horndeski theories and beyond \cite{Creminelli:2017sry,Baker:2017hug,Babichev:2016rlq,Bhattacharya:2016naa,Mukherjee:2017fqz}, look promising. In the context of $f(R)$ gravity, one adds various powers of the Ricci scalar (preferably positive) with appropriate coefficients to the Einstein-Hilbert term. The first such non-trivial correction to the Einstein-Hilbert action corresponds to the addition of a term, which is quadratic in the curvature and is believed to encode basic features of the $f(R)$ gravity \cite{star}. Even though generic higher derivative gravity theories may possibly have better UV behaviour, they are often plagued with existence of ghost modes \cite{Stelle:1976gc,Stelle:1977ry}, while this particular model with quadratic correction is ghost free for certain choice of parameters \cite{quad}.

Although the theories beyond \gr\ appear promising to address some of the well-known issues in \gr, they must also be stable under gravitational perturbation. In other words, the Hamiltonian constructed out of the dynamical degrees of freedom, associated with any gravitational theory beyond \gr, must be bounded from below. This should not only hold for perturbations around flat spacetime, but also for perturbation around de Sitter spacetime. This is because one of the main motivation of these models is to explain either inflation or late time cosmic acceleration. Thus study of stability of the dynamical degrees of freedom around maximally symmetric spacetime is of vital importance. Further, the configuration of the minima of the Hamiltonian associated with the dynamical degrees of freedom is also of interest, since any non-trivial temporal structure may lead to a time crystal-like condensate \cite{wil,will2,will3}, which has created an enormous amount of interest both in theoretical as well as in experimental physics communities (see \cite{sacha,Sacha:2017fqe} for review). 

Let us briefly point out possible effects a time crystal can have from the observational point of view and mention a few interesting recent works \cite{adda,adda1} in a similar but slightly different context. Time crystal indicates a permanent modulation in time of the lowest energy state or background configuration. Imprint of the time crystal (on the spacetime) will be felt by any field that propagates in this background. In \cite{adda} the authors have studied the Einstein–Yang-Mills equations of motion for the time evolution of the homogeneous gluon condensate in an expanding Friedman-Lemaître-Robertson-Walker (FRW, in short) cosmological background. An oscillating (in time) classical gluonic condensate field was discovered that was interpreted as a time crystal, where the continuous time-translation symmetry was broken to a discrete time-translation symmetry. A similar phenomenon occurs in our model as well, where (there are no external Yang-Mills fields and) the oscillation appears in a sector of the gravitational field due to presence of higher derivative terms in the gravitational action.

Keeping these in mind, we have used one of the simplest modification to \gr, namely, the addition of a quadratic curvature term to the gravitational action and have studied the gravitational perturbation thereof \cite{star,quad}. Subsequently, we have exploited the conventional Hamiltonian physics involving higher derivatives and have determined the structure of the minima in the momentum space. This is partially motivated by the idea of spontaneous symmetry breaking in the momentum space, suggested by one of us in \cite{sg} (also see \cite{Easson:2016klq}).Even though both the proposals of passage to the time crystal are based on condensation in momentum space through spontaneous symmetry breaking, it is worthwhile to point out the basic difference between our approach and that of \cite{will2}, leading to ground state with partial time (or, space) translation symmetry. The proposal of \cite{will2} requires at least {\it quartic} terms in field derivatives, i.e., terms of the form $\sim (\partial \phi)^4$ in the action, whereas in our approach {\it higher derivative} terms in the action play the crucial role \cite{sg}. There exist several other works exploring the idea of time crystal \cite{wil} in the cosmological context, see e.g., \cite{bain,feng,sg2,vac,sg1}. Among these works, \cite{feng} is of interest, since it has also constructed the time crystal in the gravity sector (similar to our proposal) whereas the others --- the one presented in \cite{bain} deals with the time crystal structure in the matter sector of the FRW  universe, \cite{sg2} studied time crystal in non-commutative extended gravity. In \cite{nik} it was suggested that in very strong field, such as inside a black hole, crystallization of spacetime can occur via a  phase transition. It was further shown in \cite{soda} that a subclass of Horndeski theory can be identified with a  universe having time crystal structure and in the presence of a cosmological constant, the time crystal universe can tunnel to a  de Sitter universe. Albeit in quantum systems,  space-time crystals with broken translational symmetry in time and space, was experimentally observed in interacting magnon systems \cite{trag}, in a superfluid quantum gas \cite{smit} and furthermore in \cite{kr} experiments to observe space-time crystals at room temperature in a Bose-Einstein condensate  of magnons have been proposed. Of very recent interest is the idea of spacetime crystal in a relativistic setting, suggested in \cite{gopal} and further elaborated in \cite{bojo}.

In a different perspective, for the first time, our work has revealed the form of the time crystal presented in \cite{sg},  starting from quadratic corrections to the Einstein-Hilbert action. The present work finds its basis on the following steps: (i) expanding the quadratic gravity action to the second order in the gravitational perturbation around a maximally symmetric background; (ii) starting from the Lagrangian for the perturbation and hence determining the field equations, paving way for the associated dispersion relations; (iii) constructing the Hamiltonian following \cite{ostro1,tol,and}, which requires special care, since the Lagrangian depends on higher derivative terms and finally (iv) minimizing the Hamiltonian subjected to the dispersion relation, leading to a non-trivial ground state. It needs to be stressed that no matter fields are introduced from outside and hence these non-trivial condensate-like ground states, if they exist, are induced by the spacetime geometry itself. 
 
The paper is organized as follows: In \ref{Quad_Sec} we have presented the gravitational action and its perturbation around a maximally symmetric background upto quadratic order. Using the Lagrangian derived from the perturbation, we have determined the Hamiltonian and the associated dispersion relation in \ref{Hamilton_Sec}. The idea of time crystal in non-trivial spacetime background has been discussed in \ref{time_Sec}. Subsequently, we have used the previously derived Hamiltonian, in order to determine the structure of the minima and stability of the theory in \ref{Minima_Sec}. We end in \ref{Discuss} with a discussion on the results obtained and possible future lines of research.

\emph{Notations and Conventions:} We have set the fundamental constants $c=1=\hbar$. Greek letters $\mu,\nu,\rho,\ldots$ denote spacetime indices, while Roman indices $i,j,k,\ldots$ denote spatial indices. 
\section{Quadratic action and its expansion around maximally symmetric background}\label{Quad_Sec}

In this section, we will work with the Einstein-Hilbert action with quadratic correction term, such that the gravitational action takes the following form
\begin{align}\label{action}
\mathcal{A}_{\rm grav}^{\rm quad}=\int d^{4}x\sqrt{-g}\left\{\frac{{\cal R}}{16\pi G}+\alpha {\cal R}^{2}-\frac{\Lambda}{8\pi G}\right\}~,
\end{align}
where $G$ is the Newton's constant, $\Lambda$ is the cosmological constant and $\alpha$ is a dimensionless constant that couples the higher curvature term. The gravitational field equations are obtained by varying the above action with respect to the spacetime metric $g_{\mu \nu}$, yielding,
\begin{align}
\left(1+32\pi G\alpha {\cal R}\right){\cal R}_{\mu \nu}-\frac{1}{2}\left({\cal R}+16\pi G\alpha {\cal R}^{2}\right)g_{\mu \nu}
+\left(g_{\mu \nu}\square-\nabla_{\mu}\nabla_{\nu}\right)\left\{1+32\pi G\alpha {\cal R}\right\}=8\pi G \left(-\frac{\Lambda}{8\pi G}\right)g_{\mu \nu}~.
\end{align}
Taking trace of the above equation we immediately obtain,  $-{\cal R}+3\square\left\{1+32\pi G\alpha {\cal R}\right\}=-4\Lambda$. A solution of the above equation corresponds to ${\cal R}=4\Lambda=\textrm{constant}$. Note that we have not yet specified the sign of the cosmological constant and hence the above solution encompasses both the de Sitter and the anti-de Sitter spacetime including Minkowski spacetime. Thus the background spacetime is maximally symmetric in nature.

Having described the background spacetime and the gravitational action, it is now time to consider metric perturbations around the maximally symmetric background in the context of quadratic gravity. The aim is to rewrite the gravitational Lagrangian upto quadratic order in the metric perturbation and decompose the same into transverse traceless part and an additional scalar part. The origin of this scalar can be traced back to the fact that any higher curvature theory involving Ricci scalar alone has a scalar-tensor representation. In other words, the Lagrangian written down in \ref{action} can be re-expressed as the Einstein-Hilbert action with an additional scalar field, which plays a crucial role in the stability of the theory, which we will explicitly demonstrate and is one of the main aim of this work. Following this motivation we express the metric and its inverse in terms of the perturbation $h_{\mu \nu}$ as,
\begin{align}
g_{\mu \nu}=\bar{g}_{\mu \nu}+h_{\mu \nu}~,~~g^{\mu \nu}=\bar{g}^{\mu \nu}-h^{\mu \nu}+h^{\mu \sigma}h_{\sigma}^{\nu}+\mathcal{O}(h^{3})~.
\end{align}
where $\bar{g}_{\mu \nu}$ is the background maximally symmetric spacetime, i.e., de Sitter or anti-de Sitter for the present problem, and $h_{\mu \nu}$ is the perturbation. Given the above expansion one can subsequently compute the Christoffel connection, the Riemann tensor and its descendants, i.e., Ricci tensor and Ricci scalar keeping terms quadratic in the perturbation $h_{\mu \nu}$. The detailed expression for the Ricci scalar to the quadratic order in the perturbation takes the following form,
\begin{align}\label{ricci_first}
\left(\sqrt{-g}R\right)^{(2)}&=\sqrt{-\bar{g}}\Big[\bar{R}\left(-\frac{1}{4}h_{\mu \nu}h^{\mu \nu}+\frac{1}{8}h^{2}\right)+\bar{R}_{\mu \nu}\left(h^{\mu}_{\sigma}h^{\nu \sigma}-\frac{1}{2}hh^{\mu \nu}\right)+\frac{1}{4}\bar{\nabla}_{\mu}h\bar{\nabla}^{\mu}h-\frac{1}{2}\bar{\nabla}_{\mu}h\bar{\nabla}_{\nu}h^{\mu \nu}
\nonumber
\\
&-\frac{1}{4}\bar{\nabla}_{\sigma}h^{\mu \nu}\bar{\nabla}^{\sigma}h_{\mu \nu}+\frac{1}{2}\bar{\nabla}_{\sigma}h_{\mu \nu}\bar{\nabla}^{\mu}h^{\sigma \nu}\Big]
+\textrm{Total~Derivative}~,
\end{align}
where the superscript $(2)$ denotes the fact that this action is correct upto quadratic terms of the perturbation $h_{\mu \nu}$ and $\bar R_{\mu \nu}$ is the background Ricci tensor. Further, the above expansion for the Einstein-Hilbert Lagrangian is around an arbitrary background spacetime. However, we are interested in the corresponding expression for maximally symmetric background. Thus upto quadratic order in the perturbation $h_{\mu \nu}$, the Einstein-Hilbert action with the cosmological constant term around dS/AdS background becomes \cite{quad},
\begin{align}\label{EH_Decomp}
{\mathcal{A}}_{\rm EH}^{(2)}=\frac{1}{16\pi G}\int d^{4}x\sqrt{-\bar g}\Big[\frac{1}{4}h^{\mu \nu}_{\perp}\left(\bar{\square}-\frac{\bar{R}}{6}\right)h_{\mu \nu}^{\perp}-\frac{3}{32}\Phi \left(\bar{\square}+\frac{\bar{R}}{3}\right) \Phi \Big]~,
\end{align}
where we have used the following decomposition for the perturbation $h_{\mu \nu}$ into its irreducible components, which reads \cite{quad},
\begin{align}\label{decomp}
h_{\mu \nu}=h_{\mu \nu}^{\perp}+\bar{\nabla}_{\mu}v_{\nu}+\bar{\nabla}_{\nu}v_{\mu}+\left(\bar{\nabla}_{\mu}\bar{\nabla}_{\nu}-\frac{1}{4}g_{\mu \nu}\bar{\square}\right)v+\frac{1}{4}g_{\mu \nu}h
\end{align}
where, $h_{\mu \nu}^{\perp}$ is the transverse traceless part such that $\bar{\nabla}_{\mu}h^{\mu \nu}_{\perp}=0=\bar{g}^{\mu \nu}h_{\mu \nu}^{\perp}$, $v_{i}$ is the vector part satisfying $\bar{\nabla}_{i}v^{i}=0$ and $v$ is the scalar part with $h$ being the trace of $h_{\mu \nu}$. As evident from \ref{EH_Decomp} the Einstein-Hilbert action along with the cosmological constant term when expanded upto quadratic order in the perturbation depends only on $h^{\perp}_{\mu \nu}$ as well as $v$ and $h$ through the following combination $\Phi\equiv h-\bar{\square}v$ \cite{quad}.  Note that the (background) cosmological constant is appearing in this action through the identification $\bar{R}=4\Lambda$, where $\bar{R}$ is the Ricci scalar of the background de Sitter or anti-de Sitter metric. 

We can follow an identical path for computing the contribution of the quadratic term in the gravitational Lagrangian in the perturbation. First of all, one can determine the action to quadratic order in $h_{\mu \nu}$ for an arbitrary background and then specialize to the maximally symmetric case under consideration. After a long algebra along these lines, the full action including the quadratic correction term $\sqrt{-g}R^{2}$, reduces to the following form,
\begin{align}\label{ricciquad}
\mathcal{A}_{\rm grav}^{\rm (2)}&=\int d^{4}x\sqrt{-\bar{g}}\Bigg[\frac{1}{16\pi G}\left\{\frac{1}{4}h^{\mu \nu}_{\perp}\left(\bar{\square}-\frac{\bar{R}}{6}\right)h_{\mu \nu}^{\perp} \right\} 
+\alpha \bar{R} \left\{\frac{1}{2}h^{\mu \nu}_{\perp}\left(\bar{\square}-\frac{\bar{R}}{6}\right)h_{\mu \nu}^{\perp}\right\}
\nonumber
\\
&+\frac{1}{16\pi G}\left\{-\frac{3}{32}\Phi \left(\bar{\square}+\frac{\bar{R}}{3}\right) \Phi\right\} +\alpha\left\{\frac{9}{16}\left(\bar{\square}\Phi+\frac{\bar{R}}{3}\Phi\right)^{2}
-\frac{3\bar{R}}{16}\Phi \left(\bar{\square}+\frac{\bar{R}}{3}\right)\Phi\right\}\Bigg]~.
\end{align}
It is interesting to note that the higher derivatives of the dynamical variable appear only in the $\Phi$-sector, while the evolution of the transverse and traceless sector $h^{\perp}_{\mu \nu}$ is governed by second order field equations. In particular, the Lagrangian for $h_{\mu \nu}^{\perp}$ remains identical to that in \gr, with a change in the overall factor, which now depends on both the Newton's constant $G$, as well as the higher curvature coupling $\alpha$. 

At this outset, let us discuss the physical degrees of freedom associated with this problem. As with the pure $R^{2}$ gravity model considered in \cite{quad}, here also there is a clear separation of the degrees of freedom into two parts, namely the spin-2 massless graviton $h_{\mu \nu}^{\perp}$ and the massive spin-0 scalar $\Phi$. Due to the non-flat nature of the de Sitter background, the constant Ricci scalar $\bar{R}$ of the background spacetime appears explicitly in the action, thereby making both the degrees of freedom physical as well as propagating. On the other hand, if we had considered a flat background, then a part of the scalar degree of freedom would be a ghost, which can be eliminated by using any residual gauge freedom. The existence of such a spin-0 ghost degree of freedom holds true for both Einstein gravity as well as in the presence of higher curvature corrections and is intimately tied with the flat background, see \cite{quad}. In non-flat background, e.g., the de Sitter background considered here, such ghost degrees of freedom disappears and both the spin-2 and spin-0 degrees of freedom becomes physical in $R+\alpha R^{2}$ model. While in the $\alpha \rightarrow 0$ limit, the physical, spin-0 degree of freedom becomes the ghost degree of freedom (note that kinetic parts of $h^{\perp}_{\mu \nu}$ and $\Phi$ appear with opposite signature) and can be eliminated by an appropriate choice of gauge, as pointed out in \cite{quad}.
 
The structure of the above Lagrangian suggests that the propagation of the transverse traceless mode is unaffected by the presence of the quadratic curvature term in the Lagrangian. Following which, in our subsequent analysis the transverse traceless sector chararcterized by $h^{\perp}_{\mu \nu}$ will not play any role and we will exclusively concentrate on the scalar sector $\Phi$. Extracting and combining the relevant expressions from \ref{ricciquad} associated with the scalar sector, the action for the scalar field $\Phi$, with $16\pi G \alpha \equiv \widetilde{\alpha}$, yields
\begin{align}\label{scal}
\mathcal{A}_{\rm scalar}&=\frac{1}{16\pi G}\frac{3}{16}\int d^{4}x\sqrt{-\bar{g}}\Bigg[-\left(\frac{1}{2}+\widetilde{\alpha} \bar{R}\right)\Big\{\Phi \left(\bar{\square}+\frac{\bar{R}}{3}\right)\Phi \Big\}+3\widetilde{\alpha}\left\{\left(\bar{\square}\Phi+\frac{\bar{R}}{3}\Phi\right)^{2}\right\}\Bigg]~. \end{align}
As evident from the above expression, the Lagrangian for $\Phi$ involves $\bar{\square}^{2}\Phi$, i.e., inhibits higher derivative terms originating due to inclusion of the $\alpha {\cal R}^{2}$ term in the gravitational Lagrangian. In what follows we will determine the dispersion relation associated with the above Lagrangian as well as the Hamiltonian, which we will subsequently minimize in order to determine any non-zero ground state for the scalar sector, having diverse implications. However, before delving into those details we will briefly discuss how the above action can be motivated from the scalar-tensor representation of the quadratic gravity model as well. 

\subsection{Implications from scalar-tensor theory}

Let us briefly describe the origin of the above results from the scalar-tensor representation of the quadratic gravity model. Existence of such a representation possibly originates from the Jordan frame representation of the Brans-Dicke theory \cite{Deser:1970hs,Anderson:1971dm,OHanlon:1972xqa,Wands:1993uu,Dabrowski:2005yn,Olmo:2006eh} and later on applied extensively in the context of $f(R)$ theories of gravity \cite{Sotiriou:2008rp,Nojiri:2010wj,DeFelice:2010aj}, of which the quadratic gravity model is a sub-class. The transformation to the scalar-tensor representation is achieved by a conformal transformation, in which the metric $g_{ab}$ describing the quadratic gravity model transforms to $\widetilde{g}_{\mu \nu}=\Omega^{2}g_{\mu \nu}$, where, $\Omega=(1+2\alpha \mathcal{R})^{1/2}$. Such that the gravitational action described by \ref{action}, transforms to, 
\begin{align}
\mathcal{A}_{\rm grav}^{\rm st}=\int d^{4}x\sqrt{-\widetilde{g}}\left[\frac{\widetilde{R}}{16\pi G}-\frac{1}{2}\widetilde{g}^{\mu \nu}\nabla_{\mu}\phi \nabla_{\nu}\phi-V(\phi) \right]~,
\end{align}
where, the scalar function $\phi$ and its potential $V(\phi)$ are defined as,
\begin{align}
\phi=\sqrt{\frac{3}{16\pi G}}\ln (1+2\alpha \mathcal{R})~;\qquad V(\phi)=\frac{\alpha \mathcal{R}^{2}(\phi)-2\Lambda}{16\pi G[1+2\alpha \mathcal{R}(\phi)]^{2}}~.
\end{align}
The dS background can be obtained in the scalar-tensor representation with $\widetilde{g}_{\mu \nu}$ being the dS metric and $\phi=\phi_{\rm dS}$ being a constant, such that $V(\phi_{\rm dS})$ acts as twice of the effective cosmological constant. The perturbation scheme is also straightforward, the metric is being expanded as, $\widetilde{g}_{\mu \nu}=g_{\mu \nu}^{\rm dS}+h_{\mu \nu}$ and $\phi=\phi_{\rm dS}+\widetilde{\Phi}$. Then the action will involve the perturbation of the Einstein-Hilbert action as outlined in \ref{EH_Decomp} along with contributions from the dynamic scalar perturbation $\widetilde{\Phi}$. This provides the reason why the gravitational perturbation of the quadratic gravity action cannot provide the higher derivative corrections in the transverse-traceless sector $h_{\mu \nu}^{\perp}$, since it simply comes from the Einstein term. While $\widetilde{\Phi}$, the perturbation of the scalar sector is related to the gravitational scalar sector by the relation: $\widetilde{\Phi}\propto \square \Phi$ and hence will lead to the higher derivative corrections as observed in \ref{scal}. This describes the connection between the higher curvature versus the scalar-tensor representation, with a one-to-one correspondence. 
\section{Dispersion relation and structure of the Hamiltonian}\label{Hamilton_Sec}

In the previous section, we have determined the Lagrangian associated with the scalar sector of the gravitational perturbation starting from the perturbation of the Einstein-Hilbert Lagrangian with quadratic correction. We will now vary the Lagrangian and determine the associated field equations and the associated dispersion relations by transforming to the Fourier space. Earlier, we have denoted quantities in the background spacetime with a `bar', while for notational convenience and since no confusion is likely to arise, we will remove the `bar' from all the relevant quantities of the background spacetime.

To proceed further, it is important to decompose the Lagrangian written down in \ref{scal} in terms of spatial and temporal derivatives of $\Phi$. For that purpose, it will be convenient to express the background metric in a suitable coordinate system. First of all, as emphasized earlier our primary interest is to study the early universe physics and the implications of the higher curvature term in the Lagrangian. Thus we will discuss the evolution of the scalar part of the gravitational perturbation, i.e., $\Phi$ for dS metric in the conformal time coordinate, $\eta$, defined as, $d\eta \equiv \{dt/a(t)\}$, where $t$ is the cosmic time. Therefore, the line element can be expressed in the following form,
\begin{align}
ds^{2}=a^{2}(\eta)\left(-d\eta^{2}+dx^{2}+dy^{2}+dz^{2}\right)~.
\label{aeta}
\end{align}
This can further be used to write down the Lagrangian for the scalar perturbation $\Phi$, as expressed in \ref{scal}, in terms of $\Phi'$ and the spatial derivatives of the scalar field $\Phi$ as,
\begin{align}\label{Lagrangian}
L=-\left(\frac{1}{2}+\widetilde{\alpha}R\right)a^2\Phi \left\{\nabla^2\Phi -\Phi ''-2\frac{a'}{a}\Phi ' +\frac{R}{3}a^2\Phi\right\}
+3\widetilde{\alpha} \left\{\nabla^2\Phi -\Phi ''-2\frac{a'}{a}\Phi ' +\frac{R}{3}a^2\Phi\right\}^2~.
\end{align}
Here `prime' denotes derivative with respect to the conformal time coordinate, e.g., $a'=(da/d\eta)$. Starting from the above Lagrangian it is straightforward to determine the field equation for $\Phi$, by varying the above Lagrangian with respect to $\Phi$. After discarding several total derivatives, which requires the assumption that both $\Phi$ and its time derivatives are fixed on the two end point hypersurfaces, we finally obtain, the following differential equation for $\Phi$, 
\begin{align}\label{eqn}
\left[6\widetilde{\alpha} \left\{\nabla^2-\partial_\eta ^2+2\frac{a'}{a}\partial_\eta+2\frac{d}{d\eta}\left(\frac{a'}{a}\right)\right\}-a^2\right] 
\left(\nabla^2\Phi -\Phi''-\frac{2a'}{a}\Phi'+\frac{R}{3}a^2\Phi \right)=0~.
\end{align}
As evident the field equation involves four derivatives acting on $\Phi$, which is simply a consequence of the higher derivative nature of the Lagrangian. It is also possible to derive the Hamiltonian starting from the above Lagrangian. Since the Lagrangian involves higher derivative, special care must be taken in deriving the Hamiltonian. In particular, one can employ two possible techniques for this purpose. The first one obviously corresponds to the Ostrogradsky's method \cite{ostro1,tol} of determining the Hamiltonian out of higher derivative Lagrangian. In this scheme, one treats $\Phi$ and $\Phi'$ as two independent variables and determine the momentum conjugate to each one of them. These momenta are used to eliminate $\Phi''$ from the Lagrangian and hence one determines the Hamiltonian using standard procedure, in terms of $\Phi$, $\Phi'$ and their conjugate momenta. The second procedure is due to \cite{and}, where one modifies the Lagrangian by adding some linear function of $\Phi'$ which in turn introduces a new variable, which replaces $\Phi'$. Then one determines the conjugate momentum and the associated Hamiltonian. It follows that the Hamiltonian derived by these two procedure are related by canonical transformations. Here we state the final result for the Hamiltonian, expressed in the coordinate space, which takes the following form,
\begin{align}\label{Hamiltonian}
H&=\frac{a^{2}}{2}\left[-\Phi'^{2}-(\nabla\Phi)^{2}+\frac{1}{3}Ra^{2}\Phi^{2}-2\frac{a'}{a}\Phi \Phi' \right]
+\widetilde{\alpha}\Big\{9(\Phi'')^{2}-3(\nabla^{2}\Phi)^{2}+Ra^{2} \Phi'^{2} -6\Phi''\nabla^{2}\Phi
\nonumber
\\
&+Ra^{2}(\nabla \Phi)^{2}+24\left(\frac{a'}{a}\right)^{2}\Phi'^{2}-12\frac{a''}{a}\Phi'^{2}+2Raa' \Phi \Phi' \Big\}~.
\end{align}
The terms in the first line constitutes the Hamiltonian arising out of general relativity, while the expression in the second line signifies the contribution from the higher curvature correction. As evident from the above expression, the Hamiltonian depends on $(\Phi'')^{2}$ as well as on $(\nabla^{2}\Phi)^{2}$, originating from the $R^{2}$ term in the gravitational Lagrangian. Having derived the Lagrangian as well as the Hamiltonian, we would like to transform the same in the momentum space. This will not only help us to determine the dispersion relations but also the structure of the Hamiltonian and its ground state.  

The Hamiltonian presented above must be compared with the Hamiltonians arising out of the $f(R)$ gravity as presented in \cite{Deruelle:2009pu} (see also \cite{Ohkuwa:2014mwa,Paschalidis:2011ww,Olmo:2011fh,Deruelle:2009zk,Deruelle:2007pt}). The main difference arises from the fact that, in this work we have first perturbed the Lagrangian around the de Sitter background, using the decomposition presented in \ref{decomp} and then derived the Hamiltonian by identifying the dynamical degrees of freedom. While in \cite{Deruelle:2009pu}, the Hamiltonian has been derived starting from the Lagrangian through the ADM decomposition, with no reference to perturbation around de Sitter spacetime. In this context, it is possible to construct three possible Hamiltonians, in the Einstein and Jordan frame, along with a Hamiltonian following the Ostrogradsky analysis. Among these we will briefly discuss the connection between the Hamiltonian presented here with the perturbation of the Ostrogradsky Hamiltonian in the Jordan frame, when perturbed around the de Sitter background. Using the ADM decomposition, one immediately observes from \cite{Deruelle:2009pu} that, for the $R+\alpha R^{2}$ model, there are terms $\mathcal{O}(\alpha \partial^{2}R)$. When perturbations around de Sitter background are considered, the above term will give rise to the following terms, $(\Phi'')^{2}$, $(\nabla^{2}\Phi)^{2}$ as well as the $\Phi''\nabla^{2}\Phi$ terms. Similarly, in the Ostrogradsky Hamiltonian described in \cite{Deruelle:2009pu}, there is a term $\sim \alpha\partial_{a}R\partial^{a}N$ ($N$ being the lapse function), which is responsible for the terms, $Raa'\Phi\Phi'$ as well as $a'^{2}\Phi'^{2}$. Note that the lapse function can be related to the scalar degree of freedom through the decomposition of the perturbation presented in \ref{decomp}. In this manner a direct one-to-one correspondence between the perturbed Hamiltonian of \ref{Hamiltonian} can be obtained with the perturbation of the Ostrogradsky Hamiltonian presented in \cite{Deruelle:2009pu} around the de Sitter background. However, a direct reproduction of \ref{Hamiltonian} with all the factors and signs from the results of \cite{Deruelle:2009pu} is beyond the scope of this work and will be attempted elsewhere.

However, in a time dependent situation it is difficult to write down the mode functions necessary for Fourier decomposition, which will certainly affect the present scenario as well. Thus in order to proceed further, we will choose the mode functions to be consistent with the instantaneous vacuum prescription. We will just sketch one particular way of doing the same, which is achieved by transforming to constant frequency oscillator. Briefly speaking, each Fourier mode satisfies a differential equation which resembles a harmonic oscillator with time dependent frequency and mass. Each of these harmonic oscillators can be transformed to another oscillator, but with constant mass and frequency, thus defining an instantaneous vacuum state \cite{padma,Rajeev:2017uwk}. We will assume such is the case here, such that for the scalar perturbation $\Phi$, we have the following Fourier decomposition,
\begin{align}
\Phi(\vec x,\eta)=\frac{1}{(2\pi)^2}\int d^{3}\vec k d\omega~e^{i(\vec k .\vec x -\omega \eta)}\Phi (\vec k,\omega)~.
\label{f} 
\end{align}
Since the field $\Phi$ is originating from the gravitational perturbation, it is manifestly real and hence we have $\Phi^{*}(\vec k,\omega)=\Phi(-\vec k,-\omega)$. We will assume that throughout the early evolution of the universe, the adiabaticity conditions are maintained, such that derivatives of $\omega$ and $\vec k$ can be neglected. The validity of the adiabatic approximation during the early evolution of the universe and of the scalar field will be presented in the next section. Further complications arise, as the terms involving $(\Phi'/\Phi)$ are considered, since these terms will lead to imaginary contribution to the equation of motion, which are not desirable. Interestingly all these terms are multiplied by $(a'/a)$ and hence we will also assume in accordance with the adiabaticity condition that $(a'/a)$ can be neglected. Thus substituting the above Fourier decomposition in the field equation derived from the Lagrangian, the following dispersion relations are obtained,
\begin{align}\label{d1}
{\rm dispersion~relation~I}:~~\omega^2=\vec k ^2-\frac{R}{3}a^2;\qquad 
{\rm dispersion~relation~II}:~~ \omega^2=\vec k ^2+\frac{a^2}{6\widetilde{\alpha}}~.
\end{align}
It is worth emphasizing that, even though the original Lagrangian has higher derivative terms, both the dispersion relations involve terms depending on $\omega$ and $\vec k$ in a quadratic manner. Also note that, the first dispersion relation depends on the de Sitter radius $(1/\sqrt{R})$, while the second dispersion relation depends on the coupling coefficient $\tilde \alpha$. Thus, both the de Sitter nature of the background spacetime and the presence of higher curvature term are essential to arrive at such non-trivial dispersion relations. 

Having derived the dispersion relations, let us now determine the Hamiltonian in the Fourier space, which following the expression already derived in \ref{Hamiltonian}, takes the following form,
\begin{align}\label{hh}
H(|\vec k|,\omega )&=|\Phi(\vec k,\omega)|^{2}\left\{\frac{a^{2}}{2}\left(-\omega^{2}-|\vec k|^{2}+\frac{R}{3}a^{2}\right)
+\widetilde{\alpha}\left(9\omega^{4}-3|\vec k|^{4}+a^{2}R|\vec k|^{2}+Ra^{2}\omega^{2}-6\omega^{2}|\vec k|^{2} \right) \right\}
\nonumber
\\
&\equiv f(\omega,|\vec k|)|\Phi(\omega,\vec k)|^{2}~.
\end{align}
Here the last expression defines the function $f(\omega,|\vec k|)$. Further, the first term in the Hamiltonian is originating from \gr, while the second term has its origin from the quadratic correction term in curvature. Thus we see that for each mode, characterized by $(\omega,\vec k)$, the Hamiltonian can be expressed as $f(\omega,|\vec k|)|\Phi(\omega,\vec k)|^{2}$ and it is interesting to look for the ground state by minimizing the function $f(\omega,|\vec k|)$. This is because the existence of a non-trivial ground state may require a non-zero $\widetilde{\alpha}$, which can be related to a time crystal-like behaviour and emergence of an effective cosmological constant, as we will demonstrate.
\section{Time crystal: a brief description}\label{time_Sec}

In this section, we will provide a brief description of the time crystal, in particular the classical time crystal, that is of present interest. It is well known that for conventional (spatial) crystals, spatial translation symmetry is broken and the ground state or minimum energy state requires a non-trivial spatial structure. In a similar vein, as the name suggests, time crystals refer to special type of Hamiltonian system that possesses a minimum energy state with a  non-trivial time dependence. It has already been emphasized earlier and elsewhere \cite{will2,will3,sg,feng}, emergence of time crystal structure is essentially based on spontaneous symmetry breaking in momentum space. Following the respective working principles, we would like to make a classification between the classical time crystals involved; the Hamiltonian equations of motion form \cite{will2, will3} and the Hamiltonian (energy) minimisation form \cite{sg}. Although in the present paper the latter is directly relevant for our purpose, while for completeness we briefly discuss both the forms.
\begin{itemize}

\item Hamiltonian equations of motion form of the time crystal \cite{will2}: For a generic Hamiltonian system $H(p_i,q_i)$, with $(q_i,p_i)$ being the coordinate and the conjugate momenta, respectively. The Hamiltonian equations of motion $\dot q_i=(\partial H/\partial p_i),~\dot p_i=-(\partial H/\partial q_i)$ immediately implies that for the minimum energy or, ground state, characterized by $\partial H/\partial p_i=\partial H/\partial q_i=0$, we must have $\dot q_i=0=\dot p_i$. This clearly forbids any non-trivial time dependence of the dynamical variables in the ground state and subsequently time crystals are not feasible. However, this is not the end of the story since as shown in \cite{will2}, in specially constructed models, the transition from the Lagrangian $L$ to the Hamiltonian $H$ is not smooth and in fact the Hamiltonian $H=p\dot q -L$ can become a multi-valued function of $\dot q$, thereby rendering the Hamiltonian equations of motion invalid at the minimum of the Hamiltonian $H$, where $p$ forms cusps. Typically, the minimal criteria for this to occur is when $L$ and $H$, both involve $(\dot q)^4$-term, in addition to the conventional $(\dot q)^2$-term. (For more details and explicit models see \cite{will2}).

\item Energy minimisation form of the time crystal \cite{sg}: Around the same time of the appearance of \cite{will2}, this alternative and qualitatively distinct form of the time crystal was proposed by one of the present authors \cite{sg}. This approach follows closely the conventional SSB, where the potential function $V(q)$ consists of $V(q)=Aq^4+Bq^2$ ($A,B$ being numerical parameters),  such that $V(q)$ minimizes at a non-zero (but constant) $q$. However, for a time crystal to form, the kinetic energy part $K$ requires (at least) a higher derivative quadratic term $\sim (\ddot q)^2$ together with the standard $ (\dot q)^2$-term, such that $K=C(\ddot q)^2+D(\dot q)^2$, ($C,D$ being numerical parameters). Since $q$ is a function of time alone, one can use Fourier transform to convert the same to frequency space, which is conjugate to time. Thus each time derivative of $q$ provides one factor of $\omega$, i.e., $\dot{q}\sim i\omega $, $\ddot{q}=-\omega^{2}$ and hence $(\ddot{q})^{2}\sim \omega^{4}$. For fields, on the other hand, along with time derivatives $\sim (\phi'')^{2}$, there will also be terms like $(\nabla^{2}\phi)^{2}$ (see \ref{Lagrangian}), leading to both $\omega^{4}$ and $|\vec{p}|^{4}$ term, where the components of $\vec{p}$ are conjugate to the coordinates. Clearly, in the momentum space, the kinetic energy term yields $K\sim Cp^4 + Dp^2$, which can be minimized for a non-zero $p$ in an appropriate parameter space. (Here, $p=\omega$, if it is time derivative.) This is a form of SSB albeit in momentum space leading to a minimum energy state with non-trivial time dependence; in short a time crystal. This phenomenon is demonstrated explicitly in the context of quadratic gravity model in the next section.

\end{itemize}
It is worthwhile to compare and contrast the frameworks and outcomes of the above two approaches; both of which requires an extension of the kinetic energy part, but the main difference is that a quartic $(\dot q)^4$-term is added in the former, whereas a qudratic but higher (quartic) derivative term, $(\ddot q)^2$ is added in the latter. As we note here (and explicitly show in subsequent part of the paper), this difference has far reaching consequences. First of all, higher derivative terms naturally occur in physics, and particularly so in the extended gravity theories, such as $f(R)$ gravity. Such that its linearized approximation around an appropriate background will consist of a $(\ddot q)^2$-term, as needed in the SSB time crystal. In contrast, the $(\dot q)^4$-term extension, as required in the Hamiltonian time crystal, is somewhat unnatural and ad-hoc. But more importantly, the explicit structure and time dependence of the time crystal ground state is much simpler for the SSB time crystal compared to the Hamiltonian time crystal, primarily because in SSB case, even in presence of the $(\ddot q)^2$-term, the equation of motion and dispersion relations are structurally simple with clear physical interpretation. Again these are clearly revealed in our results in the previous section, with \ref{eqn} showing the equation of motion, \ref{d1} the plane wave superposition depicting the ground state $\Phi(\vec x, \eta)$ and \ref{f} for the dispersion relations.

\section{Minimum of the Hamiltonian and an effective cosmological constant}\label{Minima_Sec}

In one of the previous sections we have worked with the higher derivative Lagrangian originating from the scalar perturbation of the Einstein-Hilbert action along with a quadratic correction. In particular, we have derived the equation of motion as well as the Hamiltonian out of the above higher derivative Lagrangian. Subsequently, we have determined the dispersion relation by transforming to the Fourier space as well as obtained the Hamiltonian in the Fourier space, which has been presented in \ref{hh}. In this section we will minimize the Hamiltonian, or to be precise we will minimize $f(\omega,|\vec k|)$ in order to determine any non-trivial ground state for the Fourier modes. It is important to realize a subtlety in the minimization procedure. Notice that $f(\omega,|\vec k|)$ depend explicitly on $a(\eta)$ which is (conformal) time dependent, see e.g., \ref{aeta}. Thus we are in fact minimizing on an arbitrary but fixed time slice. This approach was pursued in \cite{padma}, though in a different context. For this purpose, we will adopt the following procedure: (a) we will express $f(\omega,|\vec k|)$ solely in terms of either $|\vec k|$ or $\omega$ using the respective dispersion relations; (b) we will minimize $f(\omega,|\vec k|)$ in terms of either $|\vec k|$ or $\omega$ and (c) finally we shall look for any non-trivial expression for $f(\omega,|\vec k|)$ at the minimum. 

Let us start with the following dispersion relation: $\omega^{2}=|\vec k|^2-(R/3)a^2$. Then the Hamiltonian can be expressed solely in terms of either $|\vec k|$ or $\omega$, such that,
\begin{align}
f_{\rm I}(|\vec k|)&=\left(1+2\widetilde{\alpha}R\right)\left(-a^{2}|\vec k|^{2}+\frac{1}{3}Ra^{4}\right)~,
\\
f_{\rm I}(\omega)&=-\left(1+2\widetilde{\alpha}R\right)a^{2}\omega^{2}~.
\end{align}
Minimizing the first form of the function $f_{\rm I}(|\vec k|)$ with respect to $|\vec k|$, i.e., imposing $\{\partial f_{\rm I}/\partial |\vec k|\}=0$ and $\{\partial^{2} f_{\rm I}/\partial |\vec k|^{2}\}>0$, we obtain: $|\vec k|=0$ and $\omega^{2}=-(R/3)a^{2}$, along with $(1+2\widetilde{\alpha}R)<0$. However, $R$ has to be negative (for real values of $\omega$), which is not consistent with the de Sitter nature of the background spacetime. Thus this particular choice is not relevant for our discussion. On the other hand, by minimizing the function $f_{\rm I}(\omega)$ with respect to $\omega$, i.e., demanding $\{\partial f_{\rm I}/\partial \omega\}=0$ and $\{\partial^{2} f_{\rm I}/\partial \omega^{2}\}>0$ we obtain, the following conditions: $\omega=0$, $|\vec k|^{2}=(R/3)a^{2}$ and $(1+2\widetilde{\alpha}R)<0$. Note that the point where, $\widetilde{\alpha}R=-(1/2)$, denotes the point of inflection, since both $\{\partial f_{\rm I}/\partial \omega\}$ and $\{\partial^{2} f_{\rm I}/\partial \omega^{2}\}$ vanishes there. Thus at this minima we have $f_{\rm I}=0$ and hence it follows from \ref{f} that the time translation symmetry of the ground state is intact but $\Phi(\vec x)$ is periodic in $x$, indicating that the spatial translation symmetry is partially lost. However, existence of this minima either requires the coupling constant $\alpha$ or, the Ricci scalar $R$ to be negative. Among these, the scenario with negative $\alpha$ is potentially problematic due to instability issues of the parent $f(R)$ model, while the other option depicting AdS spacetime seems to be a viable option. 

On the other hand, for the other dispersion relation, i.e., $\omega^2=|\vec k|^2+(a^2/6\widetilde{\alpha})$, we obtain the following two expressions for $f(\omega,|\vec k|)$, expressed solely in terms of either $|\vec k|$ or in terms of $\omega$, as
\begin{align}
f_{\rm II}(|\vec k|)&=\left(1+2\widetilde{\alpha}R\right)\left(a^{2}|\vec k|^{2}+\frac{a^{4}}{6\widetilde{\alpha}}\right)~,
\\
f_{\rm II}(\omega)&=\left(1+2\widetilde{\alpha}R\right)a^{2}\omega^{2}~.
\end{align}
Minimizing $f_{\rm II}(\omega)$, we obtain, $\omega=0$ and $|\vec k|^{2}=-(a^{2}/6\widetilde{\alpha})$. Since $\widetilde{\alpha}>0$ is necessary for the stability of the theory, it follows that this particular minimization condition will lead to instability, as $|\vec k|^{2}$ cannot be negative and hence is discarded. On the other hand, minimizing $f_{\rm II}(|\vec k|)$ with respect to $|\vec k|$, we obtain: $|\vec k|=0$, $\omega^{2}=(a^{2}/6\widetilde{\alpha})$ along with $(1+2\widetilde{\alpha}R)>0$. Imposing these minimization conditions, we obtain the Hamiltonian to have the following structure, $H_{\rm II}=(a^{4}/6\widetilde{\alpha})(1+2\tilde{\alpha}R)|\Phi(\omega,\vec k)|^{2}$. Note that the positivity of the Hamiltonian along with the existence of a minima demands $\widetilde{\alpha}>0$, which is necessary for the stability of the gravitational action. Again, following from \ref{f} we find that the space translation symmetry of the ground state is intact but $\Phi(\eta)$ has acquired a periodicity in $\eta$ indicating that the time translation symmetry is partially lost. This is our cherished form of the time crystal. Thus we have two possible choices for the wave modes, originating from the minimization of the Hamiltonian, which read,
\begin{align}
|\vec k_{\rm I}|&=a\sqrt{\frac{R}{3}};\qquad \omega_{\rm I}=0;\qquad H_{\rm I}=0;\qquad \qquad \qquad \qquad \qquad \quad (\textrm{minima})~,
\label{min_eq_01}
\\
|\vec k_{\rm II}|&=0;\qquad \omega_{\rm II}^{2}=\frac{a^{2}}{6\widetilde{\alpha}};\qquad H_{\rm II}=\frac{a^{4}}{6\widetilde{\alpha}}(1+2\widetilde{\alpha}R)|\Phi(\omega,\vec k)|^{2};\qquad (\textrm{minima})~.
\label{min_eq_02}
\end{align}
For clarity, the above discussion about the structure of the minima of the Hamiltonian have been presented in a compact form in \ref{table:nonlin}.
\begin{table}[ht]
	\centering 
	\begin{tabular}{c c c c c} 
		\hline\hline \\
		Case & $k^2/a^2$ & $\omega^2/a^2$ & $\tilde \alpha R$  & $H/(a^4\Phi^2)$\\ [0.7ex] 
		\hline \\
		$(\omega^2,\textrm{I})$ & $R/3$ & $0$ & $<-(1/2)$ & $0$  \\ 
		$(k^2,\textrm{I})$ & $0$ & $-R/3$ & $<-(1/2)$ & $(1+2\widetilde{\alpha}R)R/3$  \\
		$(\omega^2,\textrm{II})$ & $-1/(6\widetilde{\alpha})$ & $0$ & $>-(1/2)$ & $0$  \\
		$(k^2,\textrm{II})$ & $0$ & $1/(6\widetilde\alpha)$ &  $>-(1/2)$ & $1/(6\tilde\alpha)(1+2\widetilde{\alpha}R)$  \\ [1ex] 
		\hline
	\end{tabular}
	\caption{All possible cases associated with the Minimization of the Hamiltonian.} 
	\label{table:nonlin} 
\end{table}

Having derived the structure of the minima of the Hamiltonian, as a final step we would like to see explicitly how these novel ground state solutions look like, in terms of  $\Phi(\vec x,\eta)$ given in \ref{f}. Let us first choose a simple Gaussian function, fit for a ground state, such that $\Phi(\vec k,\omega)=\exp(-\alpha k^2-\beta \omega^2)$, where $\alpha,\beta$ are two arbitrary real parameters. Considering the real part of $\Phi(\vec x,\eta)$, we obtain,
\begin{equation}\label{phi0}
\Phi(\vec x,\eta)=\frac{1}{(2\pi)^2}\int d^{3}\vec k\, d\omega~\cos(\vec k .\vec x -\omega \eta)~ \exp\left(-\alpha k^2-\beta \omega^2\right)~.
\end{equation}
Using the results from \ref{min_eq_01} and \ref{min_eq_02}, the $\Phi$ profile in the two cases reduce to
\begin{align}\label{phi1}
\Phi_{\rm I}(\vec x)&=\frac{2\pi}{(2\pi)^2}\int_0^\infty d\omega~ e^{-\beta\omega^2}~\int_0^\infty dk~ k^2 e^{-\alpha k^2}~\int_0^\pi d\theta~ \sin\theta ~ \cos(k_{\rm I} r\cos\theta) 
\nonumber
\\
&=\frac{1}{2{\sqrt{\beta\alpha^3}}}\frac{\sin(k_{\rm I}r)}{(k_{\rm I}r)}~,
\end{align}   
and,
\begin{align}\label{phi2}
\Phi_{\rm II}(\eta)&=\frac{4\pi}{(2\pi)^2}\int_0^\infty d\omega~ e^{-\beta\omega^2}~\int_0^\infty dk~ k^2 e^{-\alpha k^2}~\cos(\omega_{\rm II}\eta) 
\nonumber
\\
&=\frac{1}{2{\sqrt{\beta\alpha^3}}}~\cos(\omega_{\rm II}\eta)~.
\end{align} 
As evident, in both the cases presented above, the ground state configurations have non-trivial periodicity; in case I, $\Phi_{\rm I}(\vec x)$ has a spatial wavy character with modulated amplitude, whereas in case II, the time crystal behaviour is revealed through the periodic time varying behaviour of the ground state. From observations of wave propagation along these type of non-trivial background, one can get a handle on the cosmological parameters through $k_{\rm I}$ or $\omega_{\rm II}$.
    
Let us now comment on the validity of the adiabatic approximation in the present scenario. It follows that for the adiabatic condition to be satisfied, the Hamiltonian (which for normal mechanical systems depict the energy) should have a sufficiently large value at the minima. As \ref{table:nonlin} depicts, such is the case for the relevant choice of the dispersion relations, leading to time crystalline condensate. For which, the Hamiltonian at the minima is either of $\mathcal{O}(R)$ or of $\mathcal{O}(1/\alpha)$. Both of these are sufficiently large for adiabaticity condition to hold. 

Thus our analysis clearly demonstrates that the minimization of the Hamiltonian is intimately connected with the stability of the theory, by ensuring $\widetilde{\alpha}>0$. Hence the above provides another alternative route to assess the stability of the theory. Further, the first minimum configuration (scenario I, \ref{min_eq_01}) is associated with vanishing frequency, but non-zero wave vector and hence this depicts a situation, where there is a preferred length scale in the theory. On the other hand, in the second minimum configuration (scenario II, \ref{min_eq_02}) the modes have vanishing wave vector but a non-trivial frequency $\omega$, which is explicitly dependent on $\tilde{\alpha}$, the coupling of the quadratic correction term. In particular, this shows that the ground state of the Hamiltonian of the scalar perturbation depicts a preferred time scale in the theory $\sim \widetilde{\alpha}^{-1}$. The same trend is seen in the Hamiltonian density as well, since $(H/a^{4})\sim \widetilde{\alpha}^{-1}$. This is very much akin to the structure of the time crystal, where there is a preferred time scale in the ground state of the theory. Precisely this form of time crystal like behavior was envisaged in \cite{sg} which, however, is conceptually and structurally somewhat distinct from the ones proposed in \cite{will2}. Thus presence of higher curvature term results into scalar perturbations to behave as time crystalline state. Furthermore, the positivity of the minimum of the Hamiltonian is intimately connected with the stability of the theory. This also leads to an effective cosmological constant which depends on $\widetilde{\alpha}$. 

In addition, the above relation between the Hamiltonian of the ground state and the curvature coupling $\widetilde{\alpha}$ also enables one to provide an estimation for the higher curvature coupling term $\alpha$. This can be obtained by noting that the above scenario is most likely to appear in the context of early universe, when the universe may have experienced an inflationary expansion. During the initial stages of such an expansion, the universe had an almost de Sitter expansion and higher curvature terms are likely to be important. Thus the minimum value of the Hamiltonian can be expressed into the following form, $(\rho/M_{\rm pl}^{4})=(1/6\alpha)\{|\Phi(\omega,\vec k)|^{2}/M_{\rm pl}^{2}\}$, where we have used the result, $\widetilde{\alpha}=(\alpha/M_{\rm pl}^{2})$. From the above expression it follows that, 
\begin{align}
\alpha\sim \frac{\{|\Phi(\omega,\vec k)|^{2}/M_{\rm pl}^{2}\}}{(\rho/M_{\rm pl}^{4})}
\end{align}
During the inflationary epoch, if we want our analysis to hold, then all the energy scales must be sub-Planckian. Thus it follows that $(\rho/M_{\rm pl}^{4})\sim 1$ as well as $\{|\Phi(\omega,\vec k)|^{2}/M_{\rm pl}^{2}\}\sim 10^{-4}$, which suggests, $\alpha\sim 10^{-4}$. This is consistent with our expectation as well. Thus using inflationary dynamics, it is indeed possible to determine the coefficient of the quadratic correction $\alpha$ and thus one may impose interesting bounds on the inflationary energy scale. Therefore, the presence of higher curvature terms in the gravitational action, leads to non-trivial ground state for the Hamiltonian associated with scalar perturbation, introducing non-trivial length and time scale into the problem and an effective cosmological constant. This in turn results into an estimation of the curvature coupling through inflationary dynamics.
\section{Discussion and concluding remarks}\label{Discuss}

We have started with an action which includes the Einstein-Hilbert term as well as a term involving quadratic correction in the curvature. This being the simplest choice among the $f(R)$ theories of gravity. Such a higher curvature correction term brings in an additional length scale in the problem, namely $(16\pi G\alpha)^{-1/2}$ and is expected to modify the nature of the propagating gravitational degrees of freedom, which must be intimately connected with the stability of the corresponding gravitational theory. Following which, we have considered the ``effective" action for the gravitational perturbation, obtained by expanding the gravitational action around an arbitrary background and keeping terms quadratic in the perturbation. Surprisingly, for perturbation around a maximally symmetric background it follows that the transverse, traceless part of the gravitational perturbation satisfies a second order differential equation, even though the original action involves higher derivative terms. However, in addition to the transverse, traceless part there is also a contribution from the scalar part of the gravitational perturbations, whose evolution depends on the higher derivative terms. Thus we can conclude that, the effect of quadratic correction to the Einstein-Hilbert action is not to generate a higher derivative evolution equation for the transverse traceless part of the gravitational perturbation, rather to have an extra scalar degree of freedom, whose evolution is modified by the presence of higher derivative terms. Hence stability of the theory seemingly depends on the stability of this additional scalar degree of freedom, and following such a line of thought we have determined the existence of the minima of the Hamiltonian associated with the scalar field dynamics. 

For this purpose, we have used maximally symmetric spacetime as the background and it turns out that the stability of the Hamiltonian, i.e., the existence of a minima holds true if and only if the additional coupling parameter $\alpha$ is positive. Since for negative values of $\alpha$ the Hamiltonian becomes unbounded from below, which is a manifestation of the Ostrogradsky instability associated with higher derivative theories. Further, the structure of the minima also provides us an interesting insight about the nature of the ground state. In particular, there are two possible configurations associated with the minima of the Hamiltonian --- (a) involving non-trivial spatial wavelength $~\sqrt{(3/R)}$, associated with the cosmological length scale and (b) involving non-trivial frequency $\sim 1/(6\tilde{\alpha})$, associated with the presence of higher curvature degrees of freedom. The presence of a non-trivial temporal behaviour is reminiscent of the idea of time crystal developed in the context of many body quantum mechanical system. The above results hint at possible emergence of a time crystal-like scenario in an expanding universe with higher curvature correction. Note that the existence of time crystal-like behaviour is solely due to the presence of the higher curvature term. Moreover, the non-zero minimum value of the Hamiltonian can act as a proxy for the effective cosmological constant, since it will contribute over and above the de Sitter background. To summarize, we have demonstrated the existence of non-trivial ground state configuration for the scalar part of the gravitational perturbation in the presence of higher curvature term. Existence of such ground states are intimately related with stability of the higher curvature gravity theory and leads to non-trivial time crystal-like structure and possibly an effective cosmological constant. 

This opens up to us several future directions, in particular, existence of a time crystal-like structure in the context of quantum fields in cosmological spacetime will be a very interesting avenue to explore. Further, in this work we have discussed the quadratic correction to the Einstein-Hilbert action, thus it will be interesting to ask, whether the transverse traceless part of the gravitational perturbation remains free of higher derivatives even in the context of generic $f(R)$ theory of gravity or in higher order perturbation. Also possible connection with the scalar tensor representation can be explored. In a different perspective, it will be interesting to look at possible imprints of the new length and time scales in the propagation of gravitational waves in these non-trivial space or time dependent backgrounds. These we leave for the future.
\section*{Acknowledgements}

Research of SC is funded by the INSPIRE Faculty fellowship from Department of Science and Technology, Government of India (Reg. No. DST/INSPIRE/04/2018/000893). SC would like to thank Albert Einstein Institute, Germany for warm hospitality where a part of this work was done. 


\providecommand{\href}[2]{#2}\begingroup\raggedright\endgroup


\begin{thebibliography}{10}

\bibitem{Abbott:2016nmj}
{\bfseries Virgo, LIGO Scientific} Collaboration, B.~P. Abbott {\em et~al.},
  ``{GW151226: Observation of Gravitational Waves from a 22-Solar-Mass Binary
  Black Hole Coalescence},''
  \href{http://dx.doi.org/10.1103/PhysRevLett.116.241103}{{\em Phys. Rev.
  Lett.} {\bfseries 116} no.~24, (2016) 241103},
\href{http://arxiv.org/abs/1606.04855}{{\ttfamily arXiv:1606.04855 [gr-qc]}}.

\bibitem{TheLIGOScientific:2016src}
{\bfseries Virgo, LIGO Scientific} Collaboration, B.~P. Abbott {\em et~al.},
  ``{Tests of general relativity with GW150914},''
  \href{http://dx.doi.org/10.1103/PhysRevLett.116.221101}{{\em Phys. Rev.
  Lett.} {\bfseries 116} no.~22, (2016) 221101},
\href{http://arxiv.org/abs/1602.03841}{{\ttfamily arXiv:1602.03841 [gr-qc]}}.

\bibitem{Abbott:2017vtc}
{\bfseries VIRGO, LIGO Scientific} Collaboration, B.~P. Abbott {\em et~al.},
  ``{GW170104: Observation of a 50-Solar-Mass Binary Black Hole Coalescence at
  Redshift 0.2},'' \href{http://dx.doi.org/10.1103/PhysRevLett.118.221101}{{\em
  Phys. Rev. Lett.} {\bfseries 118} no.~22, (2017) 221101},
\href{http://arxiv.org/abs/1706.01812}{{\ttfamily arXiv:1706.01812 [gr-qc]}}.

\bibitem{Abbott:2016blz}
{\bfseries Virgo, LIGO Scientific} Collaboration, B.~P. Abbott {\em et~al.},
  ``{Observation of Gravitational Waves from a Binary Black Hole Merger},''
  \href{http://dx.doi.org/10.1103/PhysRevLett.116.061102}{{\em Phys. Rev.
  Lett.} {\bfseries 116} no.~6, (2016) 061102},
\href{http://arxiv.org/abs/1602.03837}{{\ttfamily arXiv:1602.03837 [gr-qc]}}.

\bibitem{Akiyama:2019cqa}
{\bfseries Event Horizon Telescope} Collaboration, K.~Akiyama {\em et~al.},
  ``{First M87 Event Horizon Telescope Results. I. The Shadow of the
  Supermassive Black Hole},''
  \href{http://dx.doi.org/10.3847/2041-8213/ab0ec7}{{\em Astrophys. J.}
  {\bfseries 875} no.~1, (2019) L1},
\href{http://arxiv.org/abs/1906.11238}{{\ttfamily arXiv:1906.11238
  [astro-ph.GA]}}.

\bibitem{Akiyama:2019bqs}
{\bfseries Event Horizon Telescope} Collaboration, K.~Akiyama {\em et~al.},
  ``{First M87 Event Horizon Telescope Results. IV. Imaging the Central
  Supermassive Black Hole},''
  \href{http://dx.doi.org/10.3847/2041-8213/ab0e85}{{\em Astrophys. J.}
  {\bfseries 875} no.~1, (2019) L4},
\href{http://arxiv.org/abs/1906.11241}{{\ttfamily arXiv:1906.11241
  [astro-ph.GA]}}.

\bibitem{Akiyama:2019eap}
{\bfseries Event Horizon Telescope} Collaboration, K.~Akiyama {\em et~al.},
  ``{First M87 Event Horizon Telescope Results. VI. The Shadow and Mass of the
  Central Black Hole},'' \href{http://dx.doi.org/10.3847/2041-8213/ab1141}{{\em
  Astrophys. J.} {\bfseries 875} no.~1, (2019) L6},
\href{http://arxiv.org/abs/1906.11243}{{\ttfamily arXiv:1906.11243
  [astro-ph.GA]}}.

\bibitem{Aghanim:2019ame}
{\bfseries Planck} Collaboration, N.~Aghanim {\em et~al.}, ``{Planck 2018
  results. V. CMB power spectra and likelihoods},''
\href{http://arxiv.org/abs/1907.12875}{{\ttfamily arXiv:1907.12875
  [astro-ph.CO]}}.

\bibitem{Aghanim:2018oex}
{\bfseries Planck} Collaboration, N.~Aghanim {\em et~al.}, ``{Planck 2018
  results. VIII. Gravitational lensing},''
\href{http://arxiv.org/abs/1807.06210}{{\ttfamily arXiv:1807.06210
  [astro-ph.CO]}}.

\bibitem{Clemson:2011an}
T.~Clemson, K.~Koyama, G.-B. Zhao, R.~Maartens, and J.~Valiviita,
  ``{Interacting Dark Energy -- constraints and degeneracies},''
  \href{http://dx.doi.org/10.1103/PhysRevD.85.043007}{{\em Phys. Rev.}
  {\bfseries D85} (2012) 043007},
\href{http://arxiv.org/abs/1109.6234}{{\ttfamily arXiv:1109.6234
  [astro-ph.CO]}}.

\bibitem{Lewis:2002ah}
A.~Lewis and S.~Bridle, ``{Cosmological parameters from CMB and other data: A
  Monte Carlo approach},''
  \href{http://dx.doi.org/10.1103/PhysRevD.66.103511}{{\em Phys. Rev.}
  {\bfseries D66} (2002) 103511},
\href{http://arxiv.org/abs/astro-ph/0205436}{{\ttfamily arXiv:astro-ph/0205436
  [astro-ph]}}.

\bibitem{Aghanim:2018eyx}
{\bfseries Planck} Collaboration, N.~Aghanim {\em et~al.}, ``{Planck 2018
  results. VI. Cosmological parameters},''
\href{http://arxiv.org/abs/1807.06209}{{\ttfamily arXiv:1807.06209
  [astro-ph.CO]}}.

\bibitem{Ade:2015xua}
{\bfseries Planck} Collaboration, P.~A.~R. Ade {\em et~al.}, ``{Planck 2015
  results. XIII. Cosmological parameters},''
  \href{http://dx.doi.org/10.1051/0004-6361/201525830}{{\em Astron. Astrophys.}
  {\bfseries 594} (2016) A13},
\href{http://arxiv.org/abs/1502.01589}{{\ttfamily arXiv:1502.01589
  [astro-ph.CO]}}.

\bibitem{Bamba:2012cp}
K.~Bamba, S.~Capozziello, S.~Nojiri, and S.~D. Odintsov, ``{Dark energy
  cosmology: the equivalent description via different theoretical models and
  cosmography tests},'' \href{http://dx.doi.org/10.1007/s10509-012-1181-8}{{\em
  Astrophys. Space Sci.} {\bfseries 342} (2012) 155--228},
\href{http://arxiv.org/abs/1205.3421}{{\ttfamily arXiv:1205.3421 [gr-qc]}}.

\bibitem{Copeland:2006wr}
E.~J. Copeland, M.~Sami, and S.~Tsujikawa, ``{Dynamics of dark energy},''
  \href{http://dx.doi.org/10.1142/S021827180600942X}{{\em Int. J. Mod. Phys.}
  {\bfseries D15} (2006) 1753--1936},
\href{http://arxiv.org/abs/hep-th/0603057}{{\ttfamily arXiv:hep-th/0603057
  [hep-th]}}.

\bibitem{Peebles:2002gy}
P.~J.~E. Peebles and B.~Ratra, ``{The Cosmological Constant and Dark Energy},''
  \href{http://dx.doi.org/10.1103/RevModPhys.75.559}{{\em Rev. Mod. Phys.}
  {\bfseries 75} (2003) 559--606},
  \href{http://arxiv.org/abs/astro-ph/0207347}{{\ttfamily
  arXiv:astro-ph/0207347 [astro-ph]}}.
[,592(2002)].

\bibitem{Padmanabhan:2002ji}
T.~Padmanabhan, ``{Cosmological constant: The Weight of the vacuum},''
  \href{http://dx.doi.org/10.1016/S0370-1573(03)00120-0}{{\em Phys. Rept.}
  {\bfseries 380} (2003) 235--320},
\href{http://arxiv.org/abs/hep-th/0212290}{{\ttfamily arXiv:hep-th/0212290
  [hep-th]}}.

\bibitem{Elizalde:2004mq}
E.~Elizalde, S.~Nojiri, and S.~D. Odintsov, ``{Late-time cosmology in (phantom)
  scalar-tensor theory: Dark energy and the cosmic speed-up},''
  \href{http://dx.doi.org/10.1103/PhysRevD.70.043539}{{\em Phys. Rev.}
  {\bfseries D70} (2004) 043539},
\href{http://arxiv.org/abs/hep-th/0405034}{{\ttfamily arXiv:hep-th/0405034
  [hep-th]}}.

\bibitem{Garcia-Aspeitia:2018fvw}
M.~A. Garcia-Aspeitia, A.~Hernandez-Almada, J.~Magaña, M.~H. Amante, V.~Motta,
  and C.~Martínez-Robles, ``{Brane with variable tension as a possible
  solution to the problem of the late cosmic acceleration},''
  \href{http://dx.doi.org/10.1103/PhysRevD.97.101301}{{\em Phys. Rev.}
  {\bfseries D97} no.~10, (2018) 101301},
\href{http://arxiv.org/abs/1804.05085}{{\ttfamily arXiv:1804.05085 [gr-qc]}}.

\bibitem{Bloomfield:2012ff}
J.~K. Bloomfield, E.~E. Flanagan, M.~Park, and S.~Watson, ``{Dark energy or
  modified gravity? An effective field theory approach},''
  \href{http://dx.doi.org/10.1088/1475-7516/2013/08/010}{{\em JCAP} {\bfseries
  1308} (2013) 010},
\href{http://arxiv.org/abs/1211.7054}{{\ttfamily arXiv:1211.7054
  [astro-ph.CO]}}.

\bibitem{Sotiriou:2008rp}
T.~P. Sotiriou and V.~Faraoni, ``{f(R) Theories Of Gravity},''
  \href{http://dx.doi.org/10.1103/RevModPhys.82.451}{{\em Rev. Mod. Phys.}
  {\bfseries 82} (2010) 451--497},
\href{http://arxiv.org/abs/0805.1726}{{\ttfamily arXiv:0805.1726 [gr-qc]}}.

\bibitem{Nojiri:2007as}
S.~Nojiri and S.~D. Odintsov, ``{Unifying inflation with LambdaCDM epoch in
  modified f(R) gravity consistent with Solar System tests},''
  \href{http://dx.doi.org/10.1016/j.physletb.2007.10.027}{{\em Phys. Lett.}
  {\bfseries B657} (2007) 238--245},
\href{http://arxiv.org/abs/0707.1941}{{\ttfamily arXiv:0707.1941 [hep-th]}}.

\bibitem{Nojiri:2003ft}
S.~Nojiri and S.~D. Odintsov, ``{Modified gravity with negative and positive
  powers of the curvature: Unification of the inflation and of the cosmic
  acceleration},'' \href{http://dx.doi.org/10.1103/PhysRevD.68.123512}{{\em
  Phys. Rev.} {\bfseries D68} (2003) 123512},
\href{http://arxiv.org/abs/hep-th/0307288}{{\ttfamily arXiv:hep-th/0307288
  [hep-th]}}.

\bibitem{Nojiri:2006be}
S.~Nojiri and S.~D. Odintsov, ``{Modified gravity and its reconstruction from
  the universe expansion history},''
  \href{http://dx.doi.org/10.1088/1742-6596/66/1/012005}{{\em J. Phys. Conf.
  Ser.} {\bfseries 66} (2007) 012005},
\href{http://arxiv.org/abs/hep-th/0611071}{{\ttfamily arXiv:hep-th/0611071
  [hep-th]}}.

\bibitem{DeFelice:2010aj}
A.~De~Felice and S.~Tsujikawa, ``{f(R) theories},''
  \href{http://dx.doi.org/10.12942/lrr-2010-3}{{\em Living Rev. Rel.}
  {\bfseries 13} (2010) 3},
\href{http://arxiv.org/abs/1002.4928}{{\ttfamily arXiv:1002.4928 [gr-qc]}}.

\bibitem{Nojiri:2010wj}
S.~Nojiri and S.~D. Odintsov, ``{Unified cosmic history in modified gravity:
  from F(R) theory to Lorentz non-invariant models},''
  \href{http://dx.doi.org/10.1016/j.physrep.2011.04.001}{{\em Phys. Rept.}
  {\bfseries 505} (2011) 59--144},
\href{http://arxiv.org/abs/1011.0544}{{\ttfamily arXiv:1011.0544 [gr-qc]}}.

\bibitem{Chakraborty:2016ydo}
S.~Chakraborty and S.~SenGupta, ``{Solving higher curvature gravity
  theories},'' \href{http://dx.doi.org/10.1140/epjc/s10052-016-4394-0}{{\em
  Eur. Phys. J.} {\bfseries C76} no.~10, (2016) 552},
\href{http://arxiv.org/abs/1604.05301}{{\ttfamily arXiv:1604.05301 [gr-qc]}}.

\bibitem{Chakraborty:2014xla}
S.~Chakraborty and S.~SenGupta, ``{Spherically symmetric brane spacetime with
  bulk $f(\mathcal {R})$ gravity},''
  \href{http://dx.doi.org/10.1140/epjc/s10052-014-3234-3}{{\em Eur. Phys. J.}
  {\bfseries C75} no.~1, (2015) 11},
\href{http://arxiv.org/abs/1409.4115}{{\ttfamily arXiv:1409.4115 [gr-qc]}}.

\bibitem{Chakraborty:2015bja}
S.~Chakraborty and S.~SenGupta, ``{Effective gravitational field equations on
  $m$-brane embedded in n-dimensional bulk of Einstein and $f(\mathcal {R})$
  gravity},'' \href{http://dx.doi.org/10.1140/epjc/s10052-015-3768-z}{{\em Eur.
  Phys. J.} {\bfseries C75} no.~11, (2015) 538},
\href{http://arxiv.org/abs/1504.07519}{{\ttfamily arXiv:1504.07519 [gr-qc]}}.

\bibitem{Chakraborty:2020bne}
S.~Chakraborty,
``Softly broken conformal symmetry with higher curvature terms,''
Phys. Rev. D \textbf{102} (2020) no.6, 064030
doi:10.1103/PhysRevD.102.064030
[arXiv:2004.09690 [gr-qc]].

\bibitem{Chakraborty:2015wma}
S.~Chakraborty, ``{Lanczos-Lovelock gravity from a thermodynamic
  perspective},'' \href{http://dx.doi.org/10.1007/JHEP08(2015)029}{{\em JHEP}
  {\bfseries 08} (2015) 029},
\href{http://arxiv.org/abs/1505.07272}{{\ttfamily arXiv:1505.07272 [gr-qc]}}.

\bibitem{Padmanabhan:2013xyr}
T.~Padmanabhan and D.~Kothawala, ``{Lanczos-Lovelock models of gravity},''
  \href{http://dx.doi.org/10.1016/j.physrep.2013.05.007}{{\em Phys. Rept.}
  {\bfseries 531} (2013) 115--171},
\href{http://arxiv.org/abs/1302.2151}{{\ttfamily arXiv:1302.2151 [gr-qc]}}.

\bibitem{Chakraborty:2018dvi}
S.~Chakraborty and K.~Parattu, ``{Null boundary terms for Lanczos?Lovelock
  gravity},'' \href{http://dx.doi.org/10.1007/s10714-019-2533-2,
  10.1007/s10714-019-2502-9}{{\em Gen. Rel. Grav.} {\bfseries 51} no.~2, (2019)
  23}, \href{http://arxiv.org/abs/1806.08823}{{\ttfamily arXiv:1806.08823
  [gr-qc]}}.
[Erratum: Gen. Rel. Grav.51,no.3,47(2019)].

\bibitem{Chakraborty:2017zep}
S.~Chakraborty, K.~Parattu, and T.~Padmanabhan, ``{A Novel Derivation of the
  Boundary Term for the Action in Lanczos-Lovelock Gravity},''
  \href{http://dx.doi.org/10.1007/s10714-017-2289-5}{{\em Gen. Rel. Grav.}
  {\bfseries 49} no.~9, (2017) 121},
\href{http://arxiv.org/abs/1703.00624}{{\ttfamily arXiv:1703.00624 [gr-qc]}}.

\bibitem{Chakraborty:2015taq}
S.~Chakraborty and S.~SenGupta, ``{Spherically symmetric brane in a bulk of
  $f(R)$ and Gauss?Bonnet gravity},''
  \href{http://dx.doi.org/10.1088/0264-9381/33/22/225001}{{\em Class. Quant.
  Grav.} {\bfseries 33} no.~22, (2016) 225001},
\href{http://arxiv.org/abs/1510.01953}{{\ttfamily arXiv:1510.01953 [gr-qc]}}.

\bibitem{Creminelli:2017sry}
P.~Creminelli and F.~Vernizzi, ``{Dark Energy after GW170817 and GRB170817A},''
  \href{http://dx.doi.org/10.1103/PhysRevLett.119.251302}{{\em Phys. Rev.
  Lett.} {\bfseries 119} no.~25, (2017) 251302},
\href{http://arxiv.org/abs/1710.05877}{{\ttfamily arXiv:1710.05877
  [astro-ph.CO]}}.

\bibitem{Baker:2017hug}
T.~Baker, E.~Bellini, P.~G. Ferreira, M.~Lagos, J.~Noller, and I.~Sawicki,
  ``{Strong constraints on cosmological gravity from GW170817 and GRB
  170817A},'' \href{http://dx.doi.org/10.1103/PhysRevLett.119.251301}{{\em
  Phys. Rev. Lett.} {\bfseries 119} no.~25, (2017) 251301},
\href{http://arxiv.org/abs/1710.06394}{{\ttfamily arXiv:1710.06394
  [astro-ph.CO]}}.

\bibitem{Babichev:2016rlq}
E.~Babichev, C.~Charmousis, and A.~Lehébel, ``{Black holes and stars in
  Horndeski theory},''
  \href{http://dx.doi.org/10.1088/0264-9381/33/15/154002}{{\em Class. Quant.
  Grav.} {\bfseries 33} no.~15, (2016) 154002},
\href{http://arxiv.org/abs/1604.06402}{{\ttfamily arXiv:1604.06402 [gr-qc]}}.

\bibitem{Bhattacharya:2016naa}
S.~Bhattacharya and S.~Chakraborty, ``{Constraining some Horndeski gravity
  theories},'' \href{http://dx.doi.org/10.1103/PhysRevD.95.044037}{{\em Phys.
  Rev.} {\bfseries D95} no.~4, (2017) 044037},
\href{http://arxiv.org/abs/1607.03693}{{\ttfamily arXiv:1607.03693 [gr-qc]}}.

\bibitem{Mukherjee:2017fqz}
S.~Mukherjee and S.~Chakraborty, ``{Horndeski theories confront the Gravity
  Probe B experiment},''
  \href{http://dx.doi.org/10.1103/PhysRevD.97.124007}{{\em Phys. Rev.}
  {\bfseries D97} no.~12, (2018) 124007},
\href{http://arxiv.org/abs/1712.00562}{{\ttfamily arXiv:1712.00562 [gr-qc]}}.

\bibitem{star}
A.~A. Starobinsky, ``{A New Type of Isotropic Cosmological Models Without
  Singularity},'' \href{http://dx.doi.org/10.1016/0370-2693(80)90670-X}{{\em
  Phys. Lett.} {\bfseries 91B} (1980) 99--102}.
[Adv. Ser. Astrophys. Cosmol.3,130(1987); ,771(1980)].

\bibitem{Stelle:1976gc}
K.~S. Stelle, ``{Renormalization of Higher Derivative Quantum Gravity},''
\href{http://dx.doi.org/10.1103/PhysRevD.16.953}{{\em Phys. Rev.} {\bfseries
  D16} (1977) 953--969}.

\bibitem{Stelle:1977ry}
K.~S. Stelle, ``{Classical Gravity with Higher Derivatives},''
\href{http://dx.doi.org/10.1007/BF00760427}{{\em Gen. Rel. Grav.} {\bfseries 9}
  (1978) 353--371}.

\bibitem{quad}
L.~Alvarez-Gaume, A.~Kehagias, C.~Kounnas, D.~Lüst, and A.~Riotto, ``{Aspects
  of Quadratic Gravity},'' \href{http://dx.doi.org/10.1002/prop.201500100}{{\em
  Fortsch. Phys.} {\bfseries 64} no.~2-3, (2016) 176--189},
\href{http://arxiv.org/abs/1505.07657}{{\ttfamily arXiv:1505.07657 [hep-th]}}.

\bibitem{wil}
F.~Wilczek,
``Quantum Time Crystals,''
Phys. Rev. Lett. \textbf{109} (2012), 160401
doi:10.1103/PhysRevLett.109.160401
[arXiv:1202.2539 [quant-ph]].

\bibitem{will2}
A.~Shapere and F.~Wilczek,
``Classical Time Crystals,''
Phys. Rev. Lett. \textbf{109} (2012), 160402
doi:10.1103/PhysRevLett.109.160402
[arXiv:1202.2537 [cond-mat.other]].

\bibitem{will3}
A.~Shapere and F.~Wilczek,
``Regularizations of Time Crystal Dynamics,''
Report number:	MIT-CTP/4926
[1708.03348v2 [cond-mat.stat-mech]]. 

\bibitem{sacha}
V.~Khemani, R.~Moessner, and S.~L. Sondhi, ``{A Brief History of Time
  Crystals},''
\href{http://arxiv.org/abs/1910.10745}{{\ttfamily arXiv:1910.10745
  [cond-mat.str-el]}}.

\bibitem{Sacha:2017fqe}
K.~Sacha and J.~Zakrzewski, ``{Time crystals: a review},''
  \href{http://dx.doi.org/10.1088/1361-6633/aa8b38}{{\em Rept. Prog. Phys.}
  {\bfseries 81} no.~1, (2018) 016401},
\href{http://arxiv.org/abs/1704.03735}{{\ttfamily arXiv:1704.03735
  [quant-ph]}}.

\bibitem{adda} 
A.~Addazi,  A.~Marcianò and R.~ Pasechnik, ``{Time-crystal ground state and production of gravitational
waves from QCD phase transition},''  Chinese Phys. C 43 (2019) 065101. 

\bibitem{adda1}
A.~Addazi,  A.~Marcianò,  R.~ Pasechnik and G.~Prokhorov, ``{Mirror symmetry of quantum Yang–Mills vacua and cosmological
implications},'' Eur. Phys. J. C (2019) 79:251. 

\bibitem{sg}
S.~Ghosh, ``{Emergent Discrete Space in a Generic Lifshitz Model},'' {\em
  Physica A: Statistical Mechanics and Its Applications} {\bfseries 407} (2014)
  245,
\href{http://arxiv.org/abs/1208.4438}{{\ttfamily arXiv:1208.4438 [hep-th]}}.

\bibitem{Easson:2016klq}
D.~A.~Easson and A.~Vikman,
``The Phantom of the New Oscillatory Cosmological Phase,''
[arXiv:1607.00996 [gr-qc]].

\bibitem{bain}
J.~S. Bains, M.~P. Hertzberg, and F.~Wilczek, ``{Oscillatory Attractors: A New
  Cosmological Phase},''
  \href{http://dx.doi.org/10.1088/1475-7516/2017/05/011}{{\em JCAP} {\bfseries
  1705} no.~05, (2017) 011},
\href{http://arxiv.org/abs/1512.02304}{{\ttfamily arXiv:1512.02304 [hep-th]}}.

\bibitem{feng}
X.-H. Feng, H.~Huang, S.-L. Li, H.~Lü, and H.~Wei, ``{Cosmological Time
  Crystals From Einstein-Cubic Gravities},''
\href{http://arxiv.org/abs/1807.01720}{{\ttfamily arXiv:1807.01720 [hep-th]}}.

\bibitem{sg2}
P.~Das, S.~Pan, S.~Ghosh, and P.~Pal, ``{Cosmological time crystal: Cyclic
  universe with a small cosmological constant in a toy model approach},''
  \href{http://dx.doi.org/10.1103/PhysRevD.98.024004}{{\em Phys. Rev.}
  {\bfseries D98} no.~2, (2018) 024004},
\href{http://arxiv.org/abs/1801.07970}{{\ttfamily arXiv:1801.07970 [hep-th]}}.

\bibitem{vac}
S.~I. Vacaru, ``{Space-Time Quasicrystal Structures and Inflationary and Late
  Time Evolution Dynamics in Accelerating Cosmology},''
  \href{http://dx.doi.org/10.1088/1361-6382/aaec93}{{\em Class. Quant. Grav.}
  {\bfseries 35} no.~24, (2018) 245009},
\href{http://arxiv.org/abs/1803.04810}{{\ttfamily arXiv:1803.04810
  [physics.gen-ph]}}.

\bibitem{sg1}
P.~Das, S.~Pan, and S.~Ghosh, ``{Thermodynamics and phase transition in
  Shapere-Wilczek $fgh$ model: Cosmological time crystal in quadratic
  gravity},'' \href{http://dx.doi.org/10.1016/j.physletb.2019.02.017}{{\em
  Phys. Lett.} {\bfseries B791} (2019) 66--72},
\href{http://arxiv.org/abs/1810.06606}{{\ttfamily arXiv:1810.06606 [hep-th]}}.


\bibitem {nik}
Hrvoje Nikolić,
Gravitational crystal inside the black hole, 
https://doi.org/10.1142/S0217732315502016


\bibitem{soda}
Daisuke Yoshida and Jiro Soda, 
Birth of de Sitter universe from a time crystal universe,
Phys. Rev. D 100, 123531 


\bibitem {trag}
Nick Träger , Paweł Gruszecki, Filip Lisiecki, Felix Groß, Johannes Förster, Markus Weigand, Hubert Głowiński,
Piotr Kuświk , Janusz Dubowik , Gisela Schütz, Maciej Krawczyk , and Joachim Gräfe 1,
DOI: 10.1103/PhysRevLett.126.057201


\bibitem{smit}
 J. Smits, L. Liao, H. T. C. Stoof, and P. van der Straten,
Observation of a Space-Time Crystal in a Superfluid
Quantum Gas, Phys. Rev. Lett. 121, 185301 (2018).


\bibitem{kr} A. J. E. Kreil, H. Y. Musiienko-Shmarova, S. Eggert,
A. A. Serga, B. Hillebrands, D. A. Bozhko, A. Pomyalov, and
V. S. L’vov, Tunable space-time crystal in room-temperature
magnetodielectrics, Phys. Rev. B 100, 020406(R) (2019).


\bibitem{gopal}
Venkatraman Gopalan,
Relativistic spacetime crystals. Acta Crystallographica Section A Foundations and Advances, 2021; 77 (4) DOI: 10.1107/S2053273321003259


\bibitem{bojo}
Martin Bojowald and Avadh Saxenab
From crystal color symmetry to quantum spacetime,
Acta Cryst. (2021). A77, 239-241
https://doi.org/10.1107/S2053273321005234

\bibitem{ostro1}
R.~P. Woodard, ``{Ostrogradsky's theorem on Hamiltonian instability},''
  \href{http://dx.doi.org/10.4249/scholarpedia.32243}{{\em Scholarpedia}
  {\bfseries 10} no.~8, (2015) 32243},
\href{http://arxiv.org/abs/1506.02210}{{\ttfamily arXiv:1506.02210 [hep-th]}}.

\bibitem{tol}
T.-j. Chen, M.~Fasiello, E.~A. Lim, and A.~J. Tolley, ``{Higher derivative
  theories with constraints: Exorcising Ostrogradski's Ghost},''
  \href{http://dx.doi.org/10.1088/1475-7516/2013/02/042}{{\em JCAP} {\bfseries
  1302} (2013) 042},
\href{http://arxiv.org/abs/1209.0583}{{\ttfamily arXiv:1209.0583 [hep-th]}}.

\bibitem{and}
K.~Andrzejewski, J.~Gonera, P.~Machalski, and P.~Maslanka, ``{Modified
  Hamiltonian formalism for higher-derivative theories},''
  \href{http://dx.doi.org/10.1103/PhysRevD.82.045008}{{\em Phys. Rev.}
  {\bfseries D82} (2010) 045008},
\href{http://arxiv.org/abs/1005.3941}{{\ttfamily arXiv:1005.3941 [hep-th]}}.


\bibitem{Deser:1970hs}
S.~Deser,
``Scale invariance and gravitational coupling,''
Annals Phys. \textbf{59} (1970), 248-253
doi:10.1016/0003-4916(70)90402-1



\bibitem{Anderson:1971dm}
J.~L.~Anderson,
``Scale invariance of the second kind and the brans-dicke scalar-tensor theory,''
Phys. Rev. D \textbf{3} (1971), 1689-1691
doi:10.1103/PhysRevD.3.1689



\bibitem{OHanlon:1972xqa}
J.~O'Hanlon,
``Intermediate-range gravity - a generally covariant model,''
Phys. Rev. Lett. \textbf{29} (1972), 137-138
doi:10.1103/PhysRevLett.29.137


\bibitem{Wands:1993uu}
D.~Wands,
``Extended gravity theories and the Einstein-Hilbert action,''
Class. Quant. Grav. \textbf{11} (1994), 269-280
doi:10.1088/0264-9381/11/1/025
[arXiv:gr-qc/9307034 [gr-qc]].


\bibitem{Dabrowski:2005yn}
M.~P.~Dabrowski, T.~Denkiewicz and D.~Blaschke,
``Puzzles of isotropic and anisotropic conformal cosmologies,''
Annalen Phys. \textbf{16} (2007), 237
doi:10.1002/andp.200610230
[arXiv:hep-th/0507068 [hep-th]].


\bibitem{Olmo:2006eh}
G.~J.~Olmo,
``Limit to general relativity in f(R) theories of gravity,''
Phys. Rev. D \textbf{75} (2007), 023511
doi:10.1103/PhysRevD.75.023511
[arXiv:gr-qc/0612047 [gr-qc]].


\bibitem{Deruelle:2009pu}
N.~Deruelle, Y.~Sendouda and A.~Youssef,
``Various Hamiltonian formulations of f(R) gravity and their canonical relationships,''
Phys. Rev. D \textbf{80} (2009), 084032
doi:10.1103/PhysRevD.80.084032
[arXiv:0906.4983 [gr-qc]].

\bibitem{Ohkuwa:2014mwa}
Y.~Ohkuwa and Y.~Ezawa,
``On the canonical formalism of $f(R)$-type gravity using Lie derivatives,''
Eur. Phys. J. Plus \textbf{130} (2015), 77
doi:10.1140/epjp/i2015-15077-5
[arXiv:1412.4475 [gr-qc]].

\bibitem{Paschalidis:2011ww}
V.~Paschalidis, S.~M.~H.~Halataei, S.~L.~Shapiro and I.~Sawicki,
``Constraint propagation equations of the 3+1 decomposition of f(R) gravity,''
Class. Quant. Grav. \textbf{28} (2011), 085006
doi:10.1088/0264-9381/28/8/085006
[arXiv:1103.0984 [gr-qc]].

\bibitem{Olmo:2011fh}
G.~J.~Olmo and H.~Sanchis-Alepuz,
``Hamiltonian Formulation of Palatini f(R) theories a la Brans-Dicke,''
Phys. Rev. D \textbf{83} (2011), 104036
doi:10.1103/PhysRevD.83.104036
[arXiv:1101.3403 [gr-qc]].

\bibitem{Deruelle:2009zk}
N.~Deruelle, M.~Sasaki, Y.~Sendouda and D.~Yamauchi,
``Hamiltonian formulation of f(Riemann) theories of gravity,''
Prog. Theor. Phys. \textbf{123} (2010), 169-185
doi:10.1143/PTP.123.169
[arXiv:0908.0679 [hep-th]].

\bibitem{Deruelle:2007pt}
N.~Deruelle, M.~Sasaki and Y.~Sendouda,
Prog. Theor. Phys. \textbf{119} (2008), 237-251
doi:10.1143/PTP.119.237
[arXiv:0711.1150 [gr-qc]].

\bibitem{padma}
T.~Padmanabhan, ``{Spontaneous symmetry breaking in non-inertial frames and
  curved space-time},''
\href{http://dx.doi.org/10.1016/0375-9601(82)90874-x}{{\em Phys. Lett. A}
  {\bfseries 89} (1982) 131}.

\bibitem{Rajeev:2017uwk}
K.~Rajeev, S.~Chakraborty and T.~Padmanabhan,
``Inverting a normal harmonic oscillator: physical interpretation and applications,''
Gen. Rel. Grav. \textbf{50} (2018) no.9, 116
doi:10.1007/s10714-018-2438-5
[arXiv:1712.06617 [gr-qc]].

\end{thebibliography}
\end{document}